\documentclass[twocolumn,showpacs,preprintnumbers,amsmath,amssymb]{revtex4}

\usepackage{graphicx}

\usepackage[usenames,dvipsnames]{color}

\def\ii{\' \i}

\newcommand{\beq}{\begin{equation}}
\newcommand{\eeq}{\end{equation}}
\newcommand{\beqa}{\begin{eqnarray}}
\newcommand{\eeqa}{\end{eqnarray}}
\newcommand{\bd}[1]{ \mbox{\boldmath $#1$}}

\begin{document}

\title{Phase transitions for rotational states within an algebraic cluster model}

\author{ E. L\'opez-Moreno$^{1}$, 
P. O. Hess$^{2,4}$ and H. Y\'epez-Mart\ii nez$^{3}$}

\affiliation{
$^1$ Facultad de Ciencias, Departamento de F\ii sica, UNAM. Circuito Exterior, Cd. Universitaria, Circuito Exterior, 04510 M\'exico, D.F., Mexico
}

\affiliation{
$^2$ Instituto de Ciencias Nucleares, UNAM, Circuito Exterior, C.U., A.P. 70-543, 04510 M\'exico, D.F., Mexico
}

\affiliation{
$^{3}$ Universidad Aut\'onoma de la Ciudad de M\'exico, Prolongaci\'on San Isidro 151, Col. San Lorenzo Tezonco, Del. Iztapalapa, 09790 M\'exico D.F., Mexico
}

\affiliation{
$^{4}$ Frankfurt Institute for Advanced Studies, Johann Wolfgang Goethe-Universit\"at, Ruth-Moufang-Str.1, 60438 Frankfurt am Main, Germany
}

\begin{abstract}
The {\it catastrophe theory}, an effective method for the description of
phase transitions, is applied to the {\it Semimicroscopic Algebraic
Cluster Model} (SACM) as an example of a non-trivial theory.
The ground state and excited, rotational phase transitions are 
investigated.
A short introduction to the SACM and the catastrophe theory is given.
We apply the formalism to the test case of 
$^{16}$O+$\alpha$$\rightarrow$$^{20}$Ne.
\end{abstract}
%\pacs{}
\maketitle

\section{Introduction}
{\label{sec:Introduction}}

The {\it Semimicroscopic Algebraic Cluster Model} (SACM) \cite{sacm1,sacm2} is a 
successful cluster model for nuclei. It takes into account the
Pauli exclusion principle for any number of clusters.
It was successfully applied to nuclei in the s-d shell, where one of the last
applications is published in \cite{yepez2012b}. Cluster systems of astrophysical interest were considered and also spectroscopic factors 
\cite{hess2004} calculated. 
In \cite{multichannel} different clusterizations can be related by
a {\it multichannel symmetry}, which reduces significantly the
number of independent parameters.
A detailed study of quantum phase transitions of the ground state
where published in \cite{yepez2012a,fraser2012}. However, this study was done
in a pedestrian way, not using the latest {\it technology} available, which is
the {\it catastrophe theory} \cite{gilmore1981}. Excited, rotational states were considered in \cite{morales2012}, but also here the treatment was simple.

The objective of this contribution is to apply {\it catastrophe theory} to
nuclear cluster system in order to study, in a very effective  and detailed manner, the
structure of phase transitions in nuclear cluster systems in their ground state, as well as in an excited state. The method presented here shall not only serve to understand phase 
transitions within the SACM, but also pretends to serve as an example on how
to apply catastrophe theory analysis to a non-trivial and complex model.
Thus, we hope it could be of interest also within other fields of physics.

The {\it catastrophe theory} was already applied with great success
in nuclear systems \cite{lopez1,lopez2}. 
For our study, Ref. \cite{lopez2011} shall also be of great use.

The paper is organized as follows: In section \ref{sec:SACM} 
a brief review on the SACM is presented. In section \ref{sec:Catastrophe}
an introduction to the catastrophe theory is given.  The next two sections,
\ref{sec:Application} and \ref{sec:Dneq0}, present this method applied to the particular 
cluster system. Finally, in section \ref{sec:Conclusions}
the conclusions are drawn.

\section{The Semimicroscopic Algebraic Cluster Model}
\label{sec:SACM}

The basic degrees of freedom are the relative oscillations, described by
the ${\bd \pi}^\dagger_m$ creation operators and by the corresponding
annihilation operators ${\bd \pi}^m$ ($m=\pm 1,0$). The products
of a creation with an annihilation operator form the $U_R(3)$ group,
where $R$ refers to the relative motion. A cutoff is
introduced by adding a scalar boson, whose creation and annihilation
operator are $\sigma^\dagger$ and $\sigma$ respectively. These bosons
do not have a physical meaning but are merely introduced to create a cut-off
$N$, with $N=n_\pi + n_\sigma$. The products of all types of
boson creation with boson annihilation operators form a $U_R(4)$
group.
For a detailed description of the SACM, please consult \cite{sacm1,sacm2}.

For a two-cluster system,
the model space is constructed by first calculating the product
$\left(\lambda_1,\mu_1\right)$ $\otimes$ $\left(\lambda_2,\mu_2\right)$
$\otimes$ $\left(n_\pi , 0\right)$, where the $\left(\lambda_k,\mu_k\right)$
$(k=1,2)$ describe the structure of the clusters $1$ and $2$ and $n_\pi$
is the relative oscillation number.
This results in a sum of
many $SU(3)$ irreducible representations (irreps)
$\left(\lambda , \mu \right)$ with a multiplicity
$m_{\left(\lambda , \mu \right)}$. This list has to be compared with the
content of the shell model, which can be easily obtained by computer
routines, which are available to us and can be obtained on request.
When a $SU(3)$ irrep in the former mentioned list does not appear in the
shell model, this irrep is excluded, while when it appears it is
included into
the model space of the SACM. In this manner, the model space of the SACM is
microscopic.

The Hamiltonians, however, is phenomenological and is written as a sum of
Casimir operators.
When we restrict to spherical clusters, i.e.
$\left(\lambda_k,\mu_k\right)=(0,0)$, there are
no contributions from the
individual clusters.
Two dynamical symmetry group chains of importance exist, called
the $SU_R(3)$ and the $SO_R(4)$ dynamical symmetries.
For the case discussed here, the Hamiltonian has the form

\begin{equation}
{\bf H}=x{\bf H}_{SU(3)}+(1-x){\bf H}_{SO(4)}
-\Omega' {\bd H}_{{\rm crank}}
\label{Ham}
\end{equation}
with

\beqa
{\bd H}_{SU(3)} & = & \hbar\omega{\bd n}_{\pi}+(a-b\Delta{\bd n}_{\pi})
{\bd C}_{2}({\bd n}_{\pi},0)+\xi{\bd L}^{2}
\nonumber \\
{\bd H}_{SO(4)}& = & \frac{c}{4}[({\bd \pi}^{\dagger} \cdot
{\bd \pi}^{\dagger})-( {\bd \sigma}^{\dagger} )^{2}][(
{\bf \pi}\cdot{\bd \pi} )-( {\bd \sigma} )^{2}]+\xi{\bd L}^{2}
\nonumber \\
{\bd H}_{{\rm crank}} & = & {\bd L}_{{\rm x}}
~~~,
\label{ham2}
\eeqa
where the $\Delta{\bf \hat{n}_{\pi}}={\bf \hat{n}_{\pi}}-n_{0}$ is the
excitation number of quanta and $n_{0}$ is the minimal number of relative
oscillation quanta \cite{wildermuth}, a necessary condition to satisfy the
Pauli exclusion principle.
${\bd H}_{{\rm crank}}$ is the {\it cranking} part of the Hamiltonian
\cite{eisenberg2}, where
the ${\bd L}_{{\rm x}}$ is the x-component
of the angular momentum operator and $\Omega$ is the frequency parameter
which fixes the amount of the angular momentum. The $\Omega$
is a Lagrange multiplier \cite{ring}.

The semi-classical potential is obtained by first defining a coherent
state \cite{geom}

\begin{equation}
| \alpha \rangle=\frac{N!}{(N+n_{o})!}{\cal N}_{N,n_{0}}
\frac{d^{n_{0}}}{d\gamma^{n_{0}}}
[{\bd \sigma}^\dagger + \gamma \, ({\bd \alpha}^{*}\cdot{\bd \pi}^\dagger)]^{N+n_{0}}
|0\rangle
~~~,
\end{equation}
where
\begin{equation} 
{\cal N}_{N,n_{0}}^{-2}=\frac{(N!)^{2}}{(N+n_{0})!}\frac{d^{n_{0}}}{d\gamma_{1}^{n_{0}}}\frac{d^{n_{0}}}{d\gamma_{2}^{n_{0}}}[1+\gamma_{1}\gamma_{2}(\alpha^{*}\cdot\alpha)]^{N+n_{0}} 
\label{facnor2}
\end{equation}
is the normalization constant. The $\alpha_m$ ($m=\pm 1, 0$) are the
coherent state parameters. The $\gamma_k$ are set to 1, after having applied
the derivatives. 

Finally, calculating the expectation value of the total Hamiltonian, i.e.,
$\langle \alpha \mid {\bd H} \mid \alpha \rangle$, the semi-classical
potential is obtained. When $\Omega = 0$, the only relevant parameter
is $\alpha = \alpha_0$, describing the relative distance of the two
clusters \cite{geom,yepez2012a,fraser2012}. However, in general we need all
three components when cranking is included. 

For convenience
we
use polar coordinates, i.e.

\beqa
&\alpha_{\pm1}=\frac{\alpha}{\sqrt{2}}e^{\pm i\phi}\sin(\theta)\nonumber\\
&\alpha_{0}=\alpha\cos(\theta)\
\label{condiciones}
~~~.
\eeqa

%%%%%%%%%%%%%%%%%%%%%%%%%%%%%%%%%%%%%%%%%%%%%%%%%%%%%%%%%%%%
The potential is obtained through the geometrical mapping. We add to the geometrical potential constant terms such that for $\alpha \rightarrow 0$ the potential approaches zero. The final result for the potential is \cite{yepez2012a}:

%%%%%%%%%%%%%%%%%%%%%%%%%%%%%%%%%%%%%%%%%%%%%%%%%%%%%%%%%%%%
 \begin{eqnarray}
&\widetilde{V}=A\bigg(\alpha^{2}\frac{F_{11}{(\alpha^{2})}}{F_{00}{(\alpha^{2})}}-\frac{n_{0}}{N+n_{0}}\bigg)+\nonumber\\
&  \left(B+C\sin^{2}{(2\theta)}\right) \bigg(\alpha^{4}\frac{F_{22}{(\alpha^{2})}}{F_{00}{(\alpha^{2})}}-\frac{n_{0}(n_{0}-1)}{(N+n_{0})(N+n_{0}-1)}\bigg)+\nonumber\\
&  D\, \cos{(2 \theta)}\, \alpha^{2}\, \frac{F_{20}^{N-2}{(\alpha^{2})}}{F_{00}{(\alpha^{2})}}+\nonumber\\
& \bigg(\alpha^{6}\frac{F_{33}{(\alpha^{2})}}{F_{00}{(\alpha^{2})}}-\frac{n_{0}(n_{0}-1)(n_{0}-2)}{(N+n_{0})(N+n_{0}-1)(N+n_{0}-2)}\bigg) - \nonumber\\
& \Omega(\Omega^{'})\sin{(2\theta)}\cos{(\phi)}\bigg(\alpha^{2}\frac{F_{11}{(\alpha^{2})}}{F_{00}{(\alpha^{2})}}-\frac{n_{0}}{N+n_{0}}\bigg) \, .
\label{pot-geom}
\end{eqnarray}

 The {\it Control Parameters}, $c_i = \{ A, B, C, D, \Omega \}$ are functions of the parameters in the Hamiltonian.
 
  As these control parameters varies, a complete family of analytic functions, $\widetilde{V}(\alpha, \theta, \phi: c_i )$, is generated.
 
 \begin{eqnarray}
 A(x)  =    x \, \times  \qquad  \qquad \qquad \qquad \qquad  \qquad \qquad && \nonumber \\
 \frac{(\hbar\omega+4(a-b(n_{0}-1)))+2 \xi-(1-x) \frac{c}{2} (N+n_{0}-1)}{(N+n_{0}-1)(N+n_{0}-2)(-x \, b)} && \nonumber
\end{eqnarray}

\begin{eqnarray}
B(x) & = & \frac{x\, \left(a-b\, (6-n_{0})\right) + (1-x)\, \frac{c}{2}}{(N+n_{0}-2)(-xb)}  \, ,
\nonumber \\
C(x) & = & \frac{\xi-(1-x)\, \frac{c}{4}}{(N+n_{0}-2)(-x \,b)}  \, , \nonumber \\
D(x) & = & \frac{(1-x)\, \frac{c}{2}}{(N+n_{0}-2)(-x \, b)}  \, , \label{coef}\\
\Omega(\Omega^{'},x) & = & \frac{\Omega^{'}}{(N+n_{0}-1)(N+n_{0}-2)(-x \, b)}\, .
\nonumber 
\end{eqnarray}
This structure is maintained and independent on the particular 
application to $^{16}$O+$\alpha$.

 %%%%%%%%%%%%%%%%%%%%%%%%%%%%%%%%%%%%%%%%%%%%%%%%%%%%%%%%%%%%
The potential in  Eq. (\ref{pot-geom}) is the subject of investigation in this contribution. In the next 
sections the structure under phase transitions is investigated.

\section{Catastrophe Theory and phase transitions}
\label{sec:Catastrophe}

The semi-classical behavior of the system can be studied by the methods of catastrophe theory. It is 
sufficient
to consider only the value $\phi = 0 \,  ( or \, \pi)$, and the change of variables, $v = \cos{2 \theta}$. Then the potential, Eq.(\ref{pot-geom}), can be written as
\begin{equation}
\widetilde{V} = \frac{\alpha^2}{q_0(\alpha)} \, W_0(\alpha, v; c_i) \, .
\label{pot-geom1}
\end{equation}
The $c_i = \{ A, B, C, D, \Omega \}$, are the control parameters, $q_0(\alpha)$ is a polynomial in even powers,  $q_0(0) \neq 0$ and
\begin{equation}
\begin{array}{cl}
W_0 = & \left(A \mp \Omega \,  \sqrt{1 - v^2}\right) \,  p_A(\alpha) \, + D \, v \,  p_D(\alpha)  \\
& \left(B + C \,  (1 - v^2) \right) \, p_B(\alpha) \, + p_0(\alpha) \, .
\label{pot-geom2}
\end{array}
\end{equation}
Here, for brevity, we omit the arguments into the adjustable control parameters in Eq. (\ref{coef}) and the variables $\alpha$ and $v$ are derived from those in Eq. (\ref{condiciones}).  For any values of $N$ and $n_0$ chosen, the $\{p_X(\alpha)\}$ in Eq. (\ref{pot-geom2}), results in a set of polynomials in even powers of  $\alpha$, positive integer coefficients and a non-zero constant, {\it i.e.}, $p_X(0) \neq 0$.

%%%%%%%%%%%%%%%%%%%%%%%%%%%%%%%%%%%%%%
 \begin{figure}[htbp] 
  \includegraphics[width=7.5cm,height=5cm,keepaspectratio]{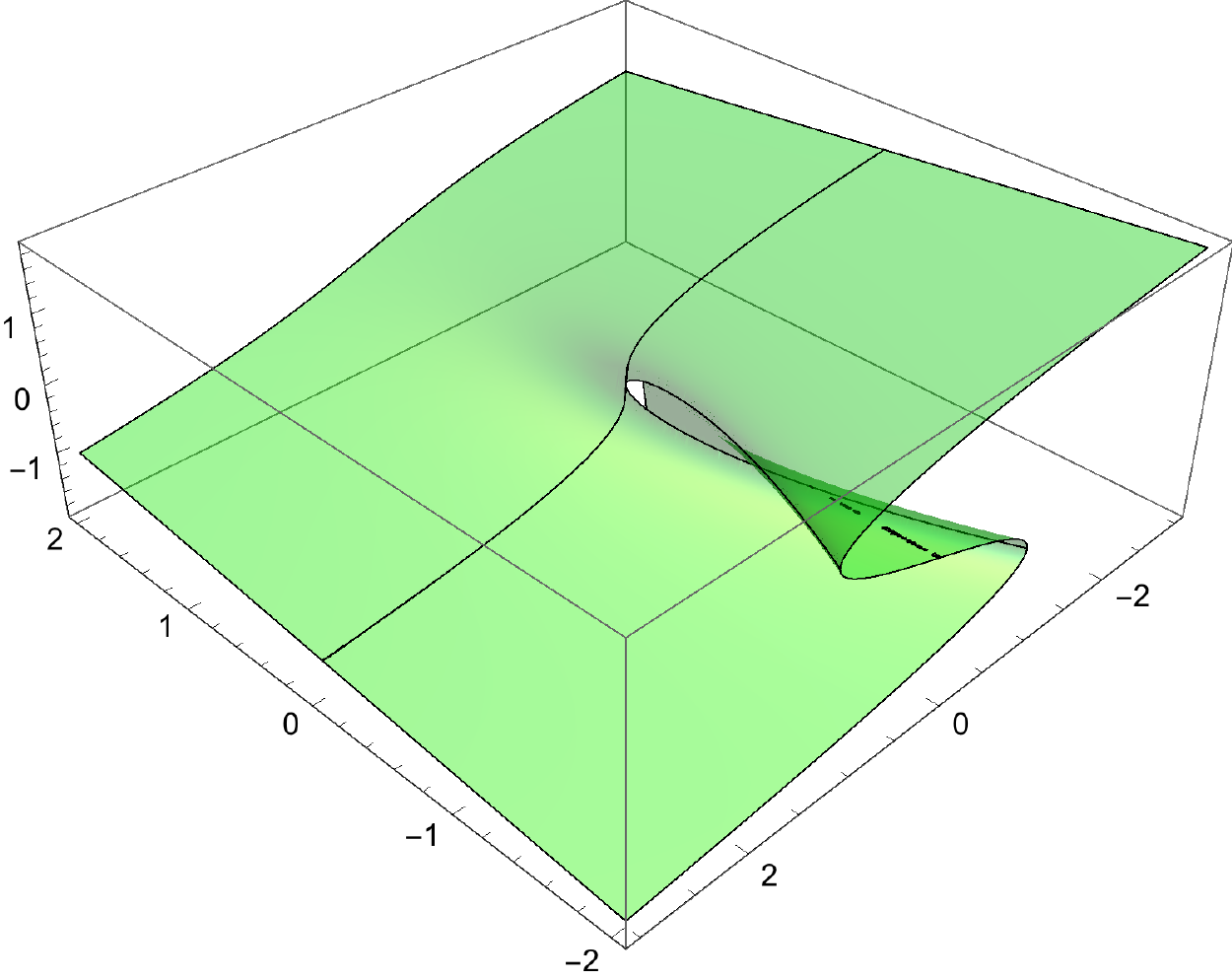} 
  \caption{The Cusp catastrophe. For the cusp potential $V_{cusp}(x)=x^4 /4+ a x^3 + b x + c$, the set of all its critical points $x_{cr}$ is given 
	by $dV_{cusp}(x; a, b) /dx = x^3 + 3\, a x^2 + b = 0$. When the essential parameters $(a,b)$ change it spans the critical manifold of the Cusp: $(a, b, x_{cr})$.}
   \label{cuspmanifold} 
  \end{figure}  
%%%%%%%%%%%%%%%%%%%%%%%%%%%%%%%%%%%%%%
%%%%%%%%%%%%%%%%%%%%%%%%%%%%%%%%%%%%%%

%%%%%%%%%%%%%%%%%%%%%%%%%%%%%%%%%%%%%%
%%%%%%%%%%%%%%%%%%%%%%%%%%%%%%%%%%%%%%
 \begin{figure}[htbp] 
  \includegraphics[width=5cm,height=7.5cm,keepaspectratio]{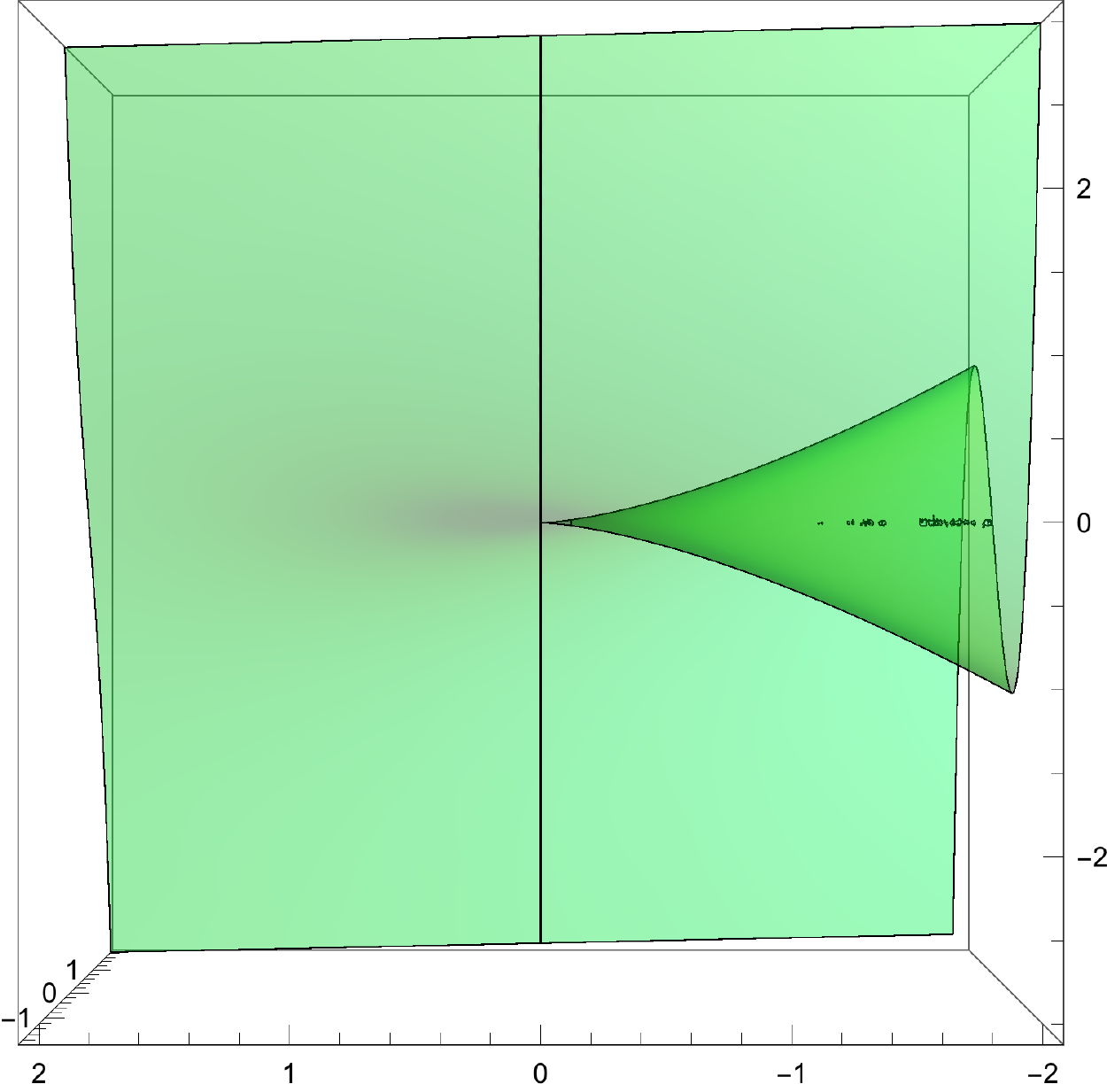} 
  \caption{Mapping the Cusp, which is a mapping singularity when the manifold in Fig. (\ref{cuspmanifold})  is projected onto the essential parameters space. The Separatrix is deduced via $(\frac{a}{3})^3 + (\frac{b}{2})^2 = 0$.}
   \label{cuspide} 
  \end{figure}  
%%%%%%%%%%%%%%%%%%%%%%%%%%%%%%%%%%%%%%
%%%%%%%%%%%%%%%%%%%%%%%%%%%%%%%%%%%%%%

In order to study the stability properties of the semi-classical system, we first evaluate the critical points of potential, $\widetilde{V}$, by means of its first derivatives: $U_\alpha \equiv \frac{\partial \widetilde{V}}{\partial \alpha}$, and $U_v \equiv \frac{\partial \widetilde{V}}{\partial v}$. We find that these can be factored as:
\begin{equation}
U_\alpha =  \frac{\alpha}{q_1(\alpha)} \, W_1(\alpha, v; c_i) \, ,
\label{alphaderivative}
\end{equation}
where, $q_1(\alpha) = \kappa_1 \, \left(\frac{q_0(\alpha)}{(1 + \alpha^2)}\right)^2$,  \ $\kappa_1$ a constant, and
\begin{equation}
U_v =  \frac{\alpha^2}{q_0(\alpha)} \, \frac{\partial W_0}{\partial v}\, .
\label{vderivative}
\end{equation}

We can see from these last expressions that the variables origin, 
$\alpha_0 = 0$, is a fundamental root: it is a critical point for the potential $\widetilde{V}$ for any choice of the control parameters. A degeneracy in any critical point means the coalescence of two or more of them due to a variation of control parameters. 
{\it In this situation, the stability of the system cannot be studied by means the usual second derivatives criteria.}
The shapes and stability of the semi-classical system can,
however,
be studied by means of the catastrophe theory program, as considered in the past \cite{lopez2}, as follows:

(i) The fundamental root, $\alpha_0 = 0$, is selected in order to obtain the {\it germ} of the potential  $\widetilde{V}$, that is, to find the point within the control parameter space where the maximum degeneracy is presented. This {\it germ} is the first term of the Taylor series expansion of $\widetilde{V}$ which can not be canceled by any set of control parameters. The first terms, in increasing order in the Taylor expansion, are successively eliminated by means of a set of relations among the control parameters. This process permits us the determination of the essential parameters, i.e., those which are a linear 
combination of the parameters used and are sufficient for the classification
of phase transitions. In Fig. (\ref{cuspmanifold}), as an example, we illustrate the critical manifold for a very simple potential. Each point in this manifold is obtained by plotting every critical point of the potential when the essential parameters vary, thus spanning a smooth surface.  

(ii) One constructs the Separatrix of the system. The first part of this separatrix consist of the bifurcation sets of the energy surface. These are the  {\it loci} within the essential parameter space where a transition occurs from one minimum to another. The bifurcation set satisfies the condition that the matrix of second derivatives of the energy surface, when evaluated at the critical points, has a null determinant value. However, in this work, in order to evaluate the bifurcation set we shall employ an important topologic feature of catastrophe theory instead \cite{gilmore1981}: the mapping, from the critical manifold into the essential parameter space, becomes singular at the bifurcation set. The critical manifold is the surface $\{(x^{(j)}_{cr}, c_i)\}$, of those points obtained when the $c_i$'s are continuously varied, for every $j$-critical point,  $(x^{(j)}_{cr}, c_i)$. The mapping mentioned becomes singular if the Jacobian of the transformation is zero, and in general the mapping is invertible except for the set points at which the tangent plane to the manifold is vertical, meaning that they are associated to the same control parameters, that is when the critical points present a coalescence.

%%%%%%%%%%%%%%%%%%%%%%%%%%%%%%%%%%%%%%
%%%%%%%%%%%%%%%%%%%%%%%%%%%%%%%%%%%%%%
\begin{figure}[t] %  figure placement: here, top, bottom, or page
   \centering
   \includegraphics[width=2in]{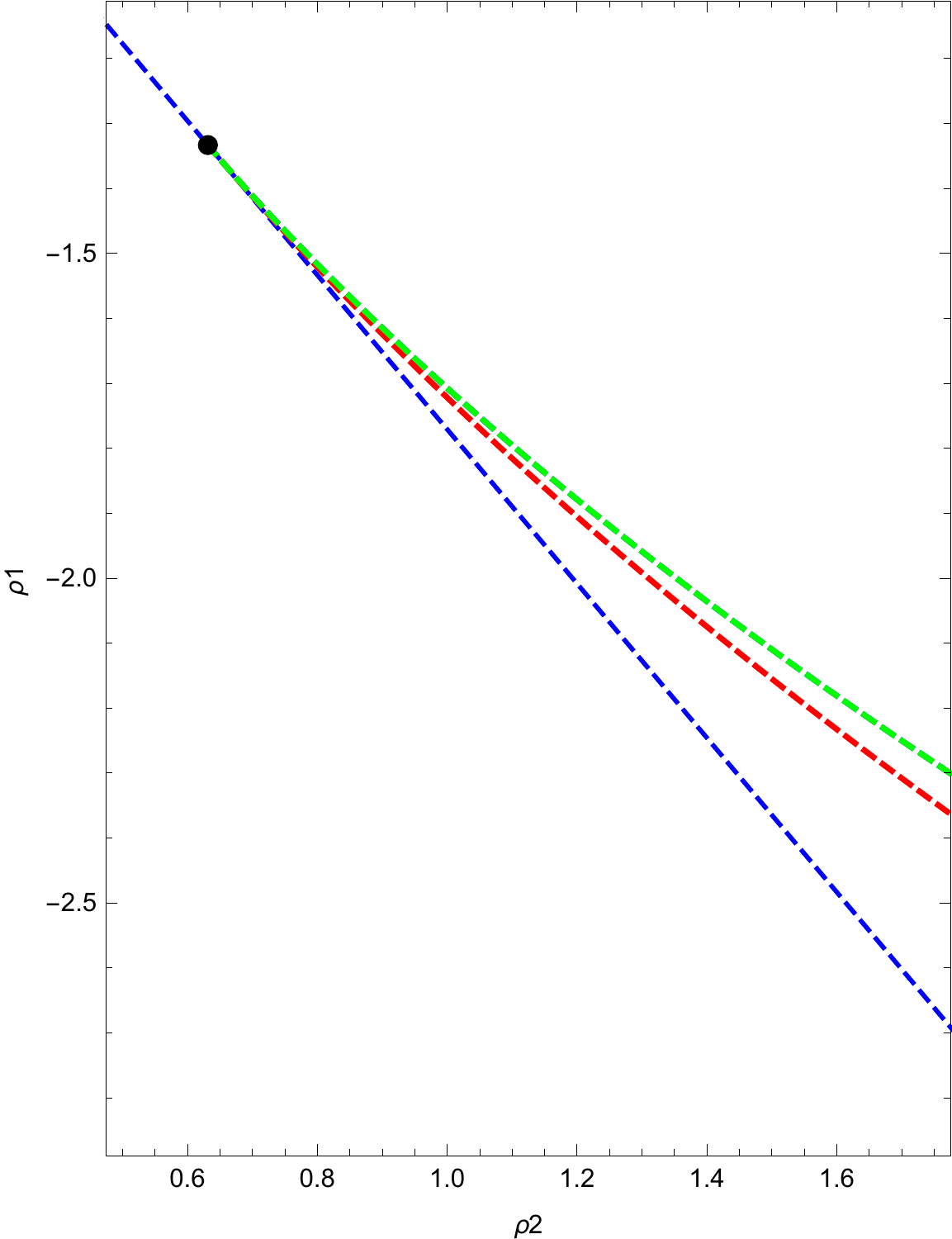} 
   \caption{The Separatrix for the $\widetilde{V}$ potential in Eq. (\ref{pot-geom}), within a subspace of the essential parameters. The straight line ${\cal C}_0$ is a bifurcation set for the fundamental root, $\alpha_0=0$ . The upper line, ${\cal C}_1$, is a bifurcation set for critical points, $\alpha_c \neq 0$. The intermediate line, ${\cal C}_M$ is a Maxwell set. The cusp, signaled by a big dot at the point (12/19, -4/3), is the locus of the {\it germ} ${\cal G}_0$. This  {\it germ} is in itself a separatrix within the subspace ${\cal C}_0$: it divides the behavior of the fundamental root, being a 
minimum 
at any point from above and going through a phase transition at this  {\it germ} it turns into 
maximum 
at the other side. }
\label{Separatrix}
\end{figure}
%%%%%%%%%%%%%%%%%%%%%%%%%%%%%%%%%%%%%
%%%%%%%%%%%%%%%%%%%%%%%%%%%%%%%%%%%%%

%%%%%%%%%%%%%%%%%%%%%%%%%%%%%%%%%%%%%%
%%%%%%%%%%%%%%%%%%%%%%%%%%%%%%%%%%%%%%
\begin{figure}[t] %  figure placement: here, top, bottom, or page
   \centering
   \includegraphics[width=2in]{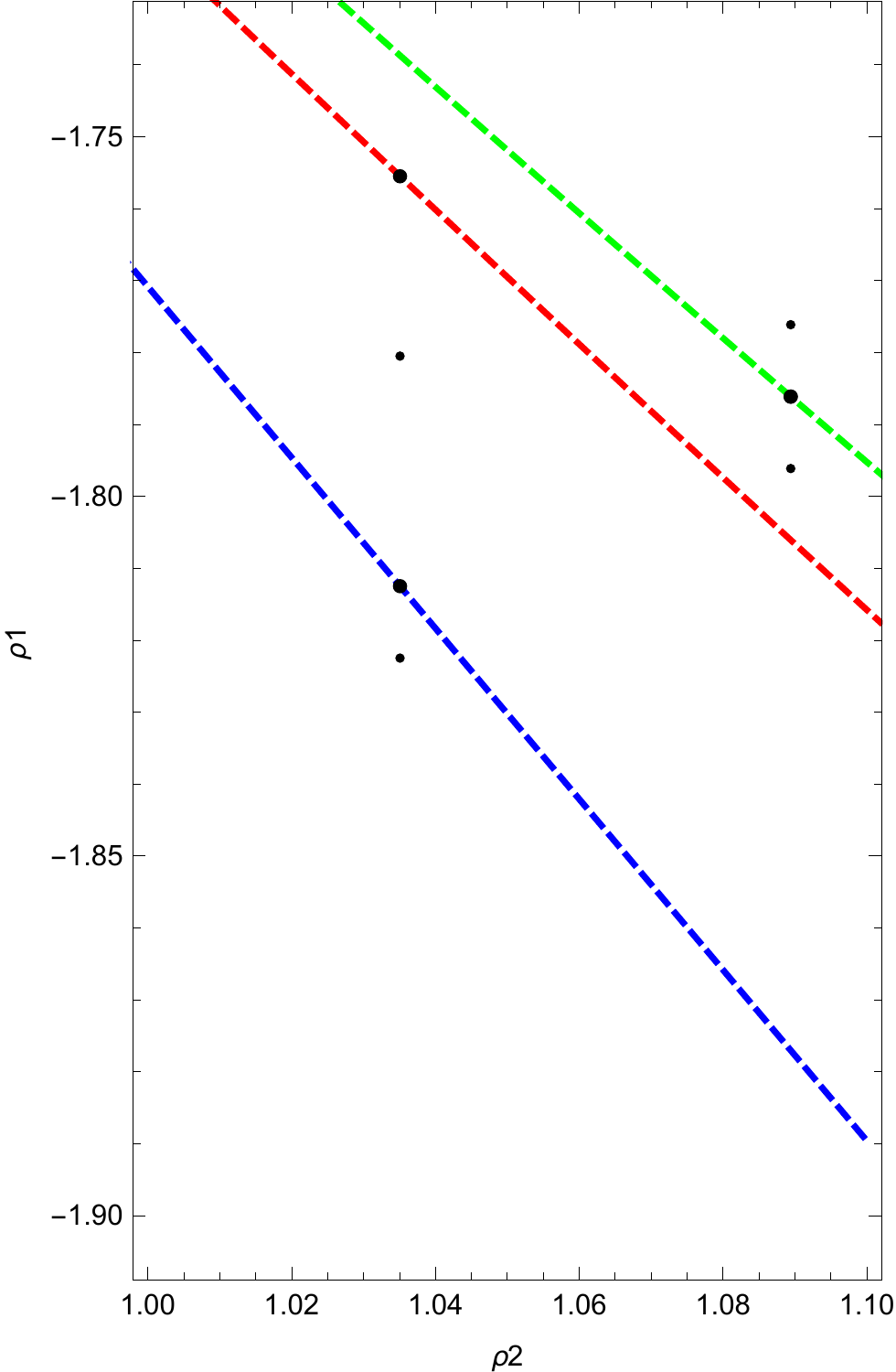} 
   \caption{A closer view of the Separatrix in Fig. (\ref{Separatrix}). Selected points at the separatrix within the subspace ${\cal D}_c$ (Eq. (\ref{separatrixD})): it divides the behavior of the fundamental root and other $\alpha_c \neq 0$ critical points, as it is shown in the next figures. }
\label{Separatriz3}
\end{figure}
%%%%%%%%%%%%%%%%%%%%%%%%%%%%%%%%%%%%%
%%%%%%%%%%%%%%%%%%%%%%%%%%%%%%%%%%%%%

(iii) The Maxwell sets are determined. These sets constitute the locus of points in the essential parameter space for which the energy surface takes the same value in two or more critical points. They can be found through the Clausius-Clapeyron equations \cite{lopez2}; however, due to the big magnitude of the problem in the present case, we shall present an original approach when solving this problem below.

(iv) Finally, the separatrix is constructed by the union of the bifurcation and the Maxwell sets. This procedure divides the essential parameter space into shape stability regions (i.e., deformed and spherical)
and identifies the loci where there are transitions according to its order. For the $\widetilde{V}$ potential in Eq. (\ref{pot-geom}), the separatrix shall be obtained, and the behavior of the model at different regions (illustrated in Figs. (\ref{Separatrix} - \ref{Separatriz3})) are explored in this paper.

%%%%%%%%%%%%%%%%%

\section{Application to the test case 
$^{16} O + \alpha \rightarrow \, ^{20}Ne$}
\label{sec:Application}

Next we follow the procedure indicated above and evaluate the numerical results for the test case of $^{16}$O+$\alpha \rightarrow$$ ^{20}$Ne. We take 
$N=12$ and $n_0=8$ as particular values, where $n_0$ is determined by the
{\it Wildermuth condition} \cite{wildermuth}.

Starting from the Taylor series expansion of the energy surface at
Eq. (\ref{pot-geom1}), around the fundamental root $\alpha_0 = 0$ ($ \phi=0$),
we obtain
\begin{equation}
\widetilde{V} = T_2 \, \alpha^2 + T_3 \, v \; \alpha^2 + 
T_4 \, \alpha^4 + T_5 \, v \, \alpha^4 + T_6 \, \alpha^6 + \cdots \, ,
\label{TaylorV}
\end{equation}
where
\begin{equation}
\begin{array}{cl}
T_2 = & 18\,  (28 + 57\,  \rho_2 + 48 \,  \rho_1)/95 \, ,\\
T_3 = & 594 \,  D/19 \, , \\
T_4 = & 18\,  (28 + 57 \, \rho_2 + 48 \,  \rho_1)/95 \, , \\
T_5 = &-508464\,  D/19 \, , \\
T_6 = & 144\,  (809893 + 1930077\,  \rho_2 + 1519398 \, \rho_1)/19 \, , \\
\rho_2 \equiv &(A - \Omega) \\
\rho_1 \equiv &(B  + C) \, .
\label{TaylorTs}
\end{array}
\end{equation}
The first order term ($\sim \alpha$) in Eq.(\ref{TaylorV}) is obviously eliminated because it is evaluated at a critical point. 

The $T_2$ term is eliminated at the points of the following hyper surface  in the parameter space
\begin{equation} 
{\cal C}_0 = \left\{(\rho_2, \rho_{1c}) \, \Bigg\vert  \,  \rho_{1c}(\rho_2) = -\frac{19}{16}\, \rho_2 - \frac{7}{12} \right\} \, .
\label{separatrix0}
\end{equation}
At ${\cal C}_0$ the fundamental root becomes double degenerate because their second derivatives are canceled. 

The $T_3$ term cancels and thus the fundamental root will degenerate further 
for those points in parameter space laying on the plane ${\cal D}_c $, defined by:
\begin{equation}
D \rightarrow D_c = 0 \, .
\label{separatrixD}
\end{equation}
On this plane, variable-$v$ becomes double-degenerated, as we shall demonstrate later. It is important to observe that the plane ${\cal D}_c $ corresponds to the limit $x = 1$, as it is clear from Eq.(\ref{coef}), 
it corresponds to $SU(3)$-limit in Eq.(\ref{Ham}).

The $T_4$ term, corresponding to the fourth order derivatives evaluated at the fundamental root,  is eliminated when 
Eqs. (\ref{separatrix0}) and (\ref{separatrixD}) are satisfied, and simultaneously the following values are chosen:
\begin{equation} 
(\rho_2, \rho_{1c}(12/19))= ( 12/19,-4/3) \, ,
\label{germen0}
\end{equation}
thus increasing to the fourth order the $\alpha$ degeneracy. Under these conditions, the restriction 
\begin{equation}
 \Omega \rightarrow \Omega_c \equiv \frac{32}{19} \, C \, ,
 \label{germen1}
 \end{equation}
shall further increase the degeneracy of the $v$-variable to fourth order, being $v_0=0$ the unique variable-$v$ critical point, as we shall show later.

The term $T_5$ will be then automatically canceled.

\begin{figure}[t] %  figure placement: here, top, bottom, or page
   \centering
   \includegraphics[width=3in]{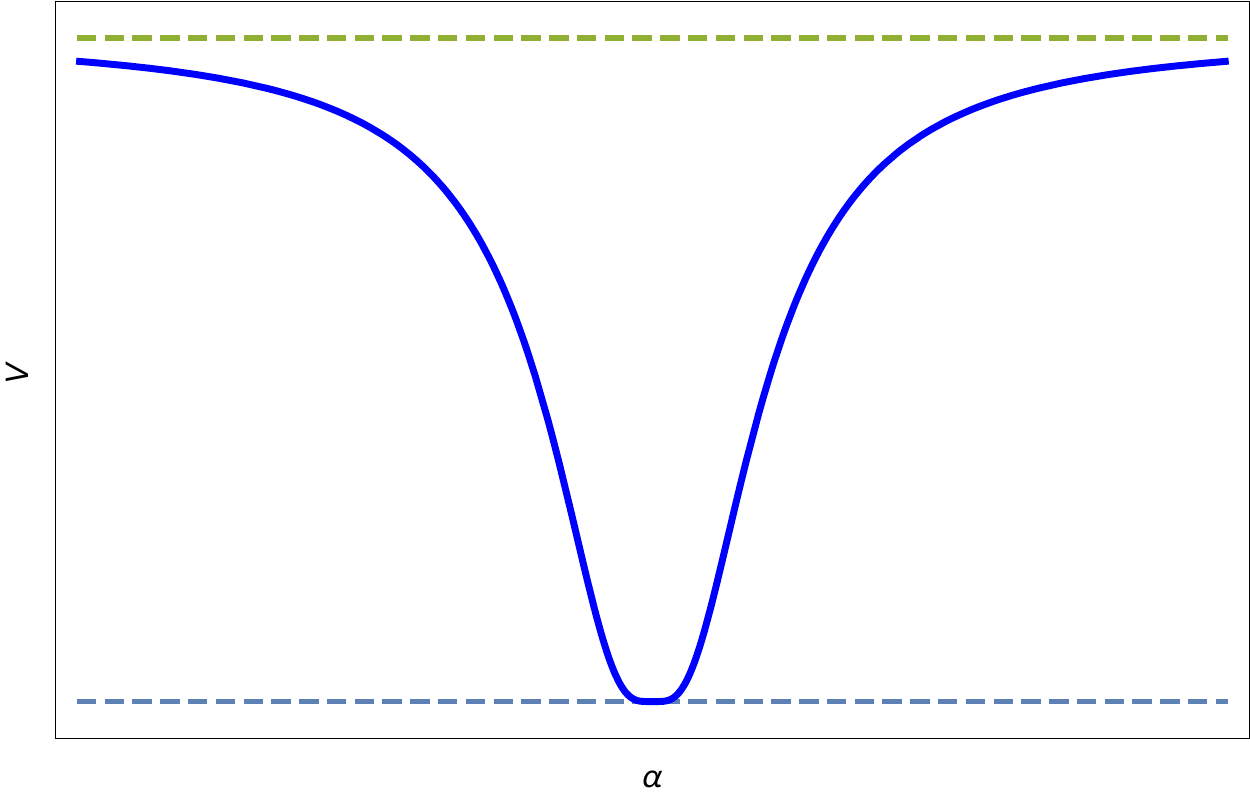} 
   \caption{Energy Surface $\widetilde{V}(\alpha, v)$, evaluated at the critical {\it germ} in the essential parameter space.  The minimum at the origin cannot be approximated by an harmonic oscillator. It has an asymptotic value, $\widetilde{V}(\alpha \rightarrow \infty, v=0)$, indicated as the upper horizontal line.}
   \label{GermV}
\end{figure}

It is important to note that when conditions, Eq.
(\ref{separatrix0}) to Eq.
(\ref{germen1}) are satisfied, it is impossible to cancel the sixth order term in Taylor series expansion by any further parameter relation. In this way, we  obtain the {\it germ}  ${\cal G}_0$ of the energy potential;  it corresponds to the point within the parameter space satisfying all the above conditions:
\begin{equation}
{\cal G}_0\;: (\rho_{2G},\, \rho_{1G},\, D_c) = (12/19,\,-4/3,\, 0) \, .
\label{Germ0}
\end{equation}
When evaluated at this  {\it germ} in the parameter space, the potential becomes: a) six-fold degenerated at its minimum, b) parameter independent, and c) a ratio of two polynomials,  $\widetilde{V}(\alpha, v=0)= {\cal P}_n(\alpha)/{\cal P}_d(\alpha)$; with, ${\cal P}_n(0)=0$, ${\cal P}_d(0) \neq 0$, and  $\widetilde{V}(\alpha \rightarrow \infty, v) = 11/57$ (see Fig.(\ref{GermV})). At its  {\it germ} point, the potential $\widetilde{V}$ has no other minimal critical point, and its maximum is attained only at infinity. If any of the parameters vary, going out of this {\it germ}, then other shapes and phase-shapes-transitions can occur. 
%%%%%%%%%%%%%%%%%%%%%%%%%%%%%%%%%%%%%%
%%%%%%%%%%%%%%%%%%%%%%%%%%%%%%%%%%%%%%
%%%%%%%%%%%%%%%%%%%%%%%%%%%%%%%%%%%%%
%%%%%%%%%%%%%%%%%%%%%%%%%%%%%%%%%%%%%

From the above discussion we found the {\it germ}, ${\cal G}_0$, and conclude that the separatrix for the fundamental root consist of the plane ${\cal D}_c $, Eq. (\ref{separatrixD}), and the straight line 
${\cal C}_0$, Eq (\ref{separatrix0}). On the one hand, out of separatrix  ${\cal D}_c $, as it will be shown later, the doubly degenerate $v_0=0$ critical point  bifurcates and $v = 0$ is no longer an extreme value; this implies a shape transition in variable $v$-direction for the fundamental root. On the other hand, at the separatrix  line  ${\cal C}_0$ within the plane ${\cal D}_c $ there is a shape transition in the variable $\alpha$. By crossing this line 
from above, a minimum changes to a maximum. If one goes through a continuous change within the line  ${\cal C}_0$, then  there is a shape transition in the variable $\alpha$. Crossing the {\it germ} in one direction (from left): A minimum changes to a maximum.

%%%%%%%%%%%%%%%%%%%%%%%%%%%%%%%%%%%%%%
%%%%%%%%%%%%%%%%%%%%%%%%%%%%%%%%%%%%%%
\begin{figure}[t] %  figure placement: here, top, bottom, or page
   \centering
   \includegraphics[width=2in]{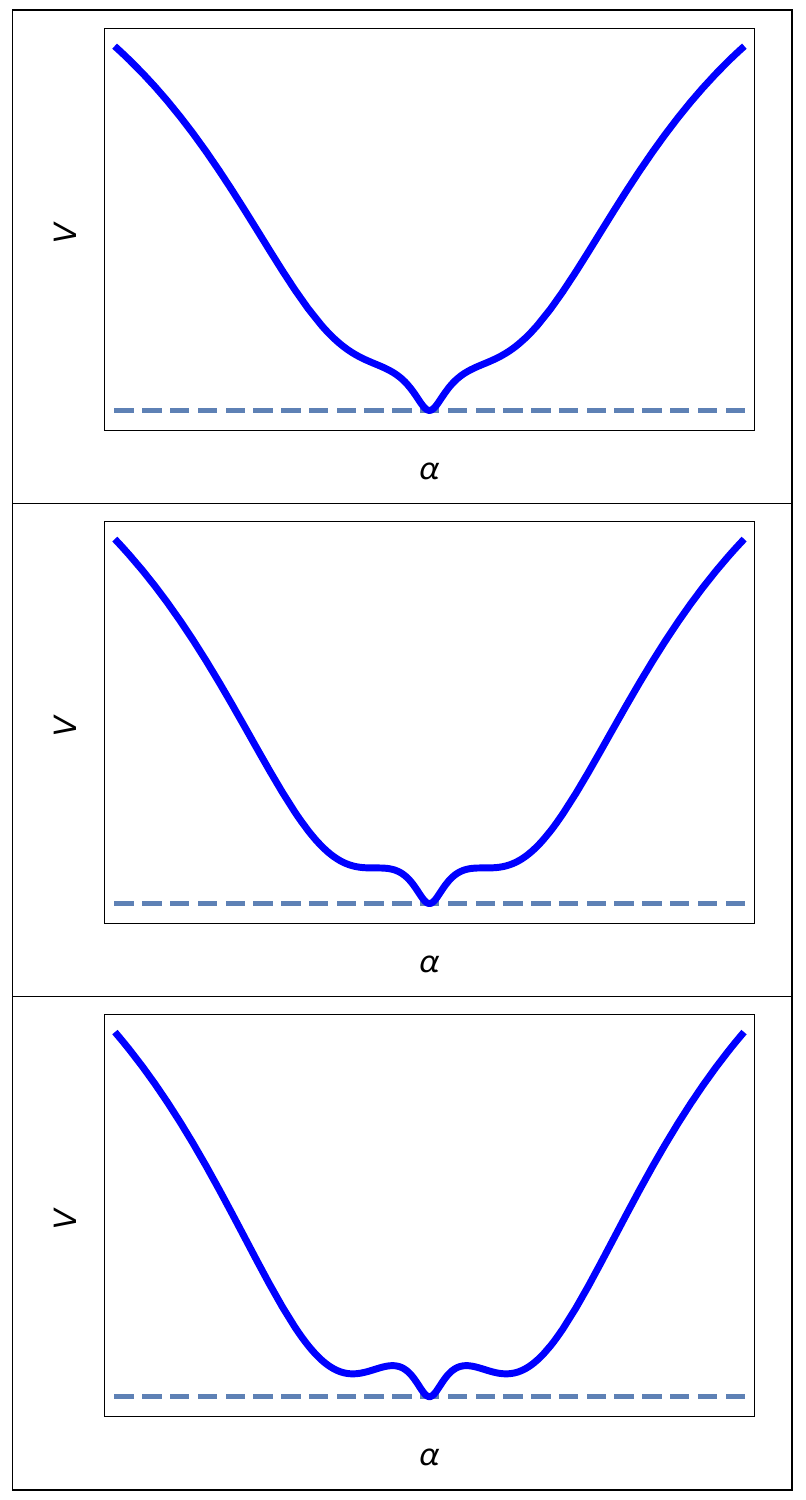} 
   \caption{Shape transition at bifurcation set ${\cal C}_1$, at the parameter points indicated in Fig. (\ref{Separatriz3}). 
   The top figure shows a minimum at the $\alpha=0$ coordinates origin. This is the general behavior at the upper region, above the  ${\cal C}_1$ set. The middle figure corresponds to a point on this set; there, a critical degenerated saddle point appears. The bottom figure shows the behavior of the potential at the region in between the ${\cal C}_1$, and the ${\cal C}_M$ sets: a metastable $\alpha \neq 0$ appears.}
\label{Grida}
\end{figure}
%%%%%%%%%%%%%%%%%%%%%%%%%%%%%%%%%%%%%
%%%%%%%%%%%%%%%%%%%%%%%%%%%%%%%%%%%%%

%%%%%%%%%%%%%%%%%%%%%%%%%%%%%%%%%%%%%%
%%%%%%%%%%%%%%%%%%%%%%%%%%%%%%%%%%%%%%
\begin{figure}[t] %  figure placement: here, top, bottom, or page
   \centering
   \includegraphics[width=2in]{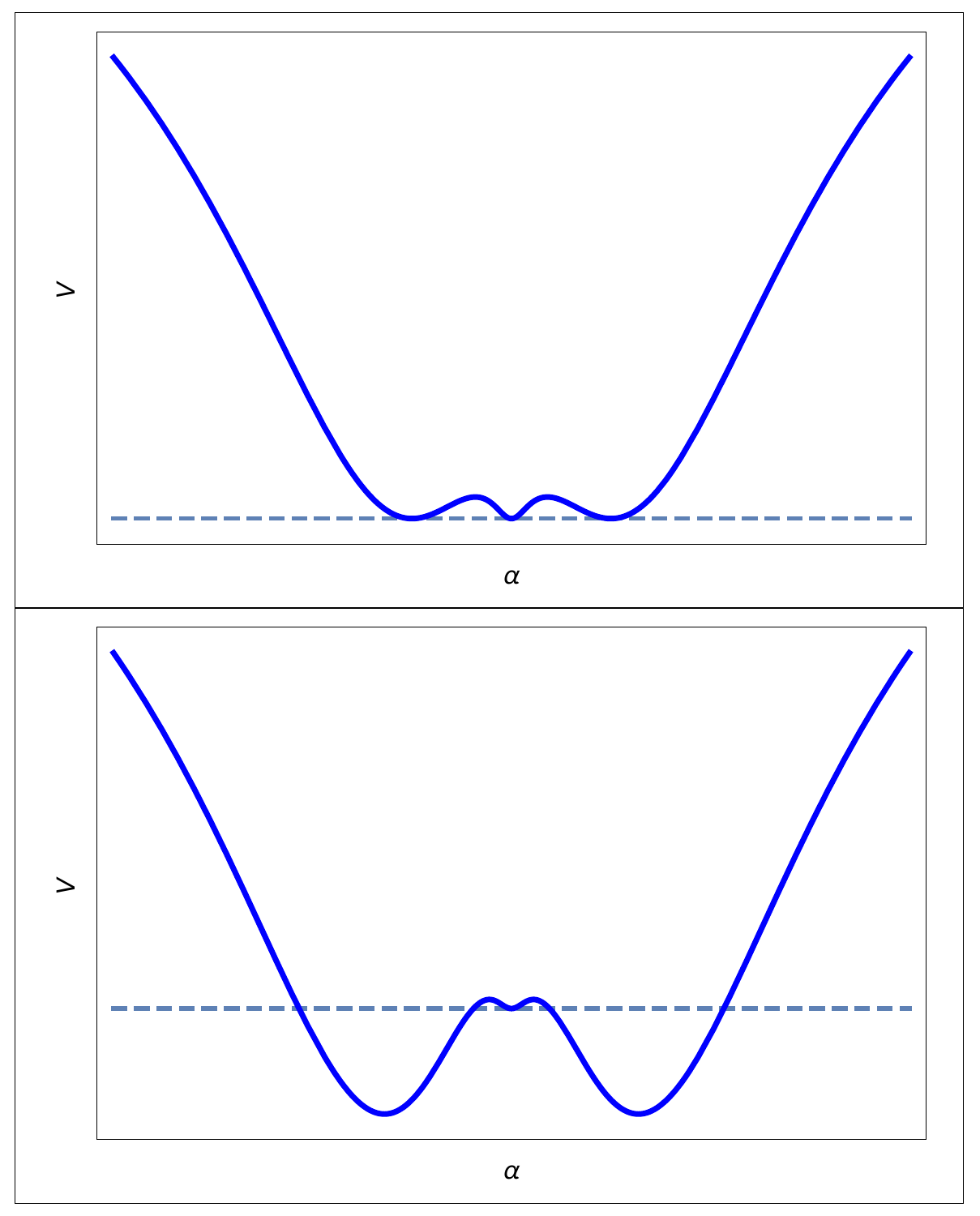} 
   \caption{Shape transition at the Maxwell set ${\cal C}_M$, at the parameter points indicated in Fig. (\ref{Separatriz3}). 
   The top figure exhibits the coexistence of two minima at the Maxwell set. At the bottom figure, the potential is evaluated at the point between the ${\cal C}_M$ and the ${\cal C}_0$ sets. There, the fundamental root corresponds to an excited metastable state, and the ground state is an $\alpha \neq 0$ deformed state. }
\label{Gridb}
\end{figure}
%%%%%%%%%%%%%%%%%%%%%%%%%%%%%%%%%%%%%
%%%%%%%%%%%%%%%%%%%%%%%%%%%%%%%%%%%%%%

%%%%%%%%%%%%%%%%%%%%%%%%%%%%%%%%%%%%%
%%%%%%%%%%%%%%%%%%%%%%%%%%%%%%%%%%%%%%
\begin{figure}[t] %  figure placement: here, top, bottom, or page
   \centering
   \includegraphics[width=2in]{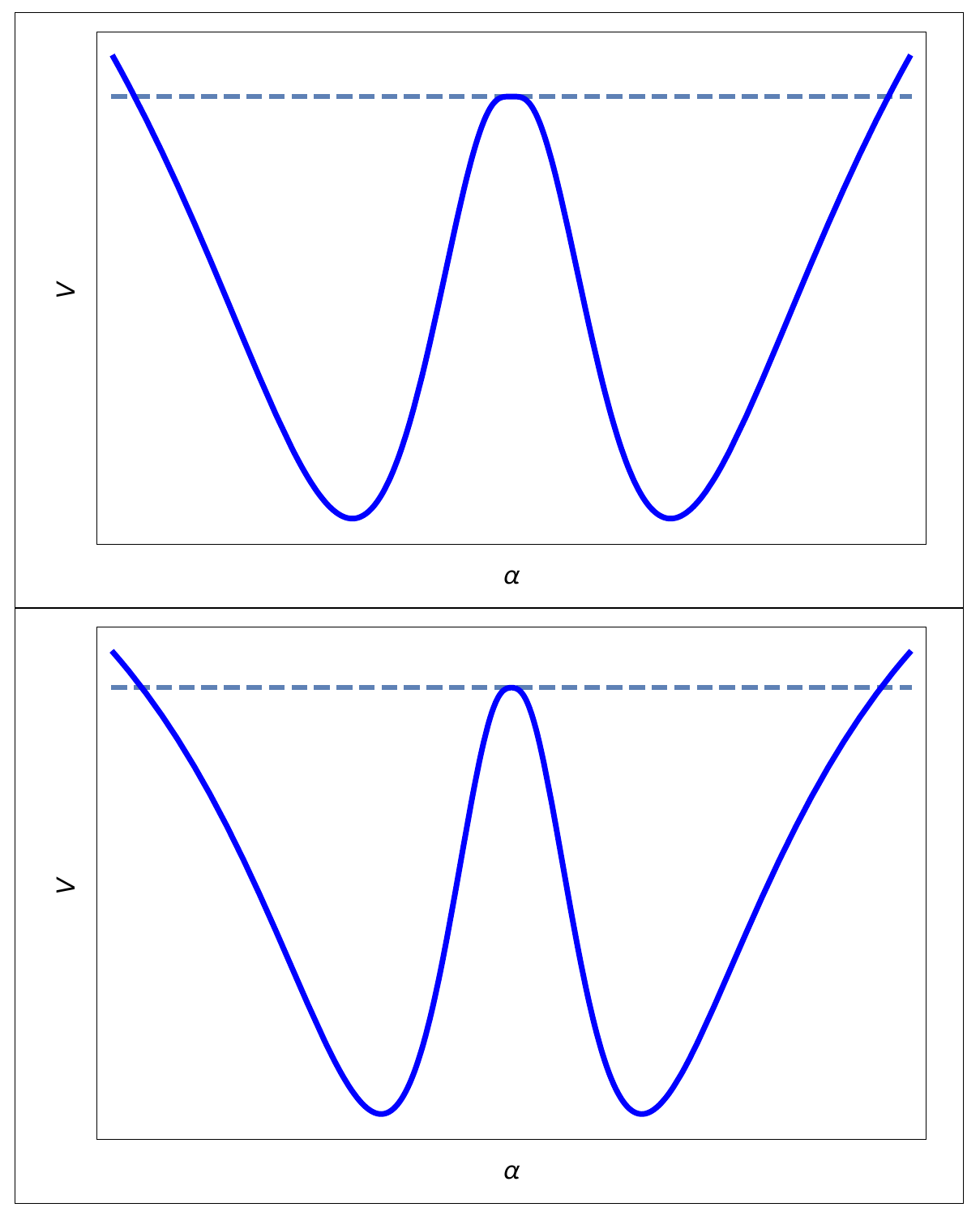} 
   \caption{Shape transition at  the ${\cal C}_0$ set in 
	Fig. (\ref{Separatriz3}). The top figure exhibits degenerated maximum at the origin and a deformed minimum at $\alpha_c \neq 0$. The bottom figure shows the behavior for all those points bellow the ${\cal C}_0$ set: The potential has only one $\alpha_c \neq 0$ minimum and a regular maximum at the origin.}
\label{Gridc}
\end{figure}
%%%%%%%%%%%%%%%%%%%%%%%%%%%%%%%%%%%%%
%%%%%%%%%%%%%%%%%%%%%%%%%%%%%%%%%%%%%

In this contribution we shall be interested in points very near to the separatrix plane ${\cal D}_c $, so we consider the parameter values $D \sim 0$; however, we shall study the more general bifurcation and the Maxwell sets in the variable $\alpha$, for all $\alpha$-extreme values with $v_{cr} = 0$. At the last part of this work, a very brief numerical example of bifurcation in the $v$-critical point shows us that a spontaneous symmetry breaking phenomena is represented as a phase transition and also we present a numerical example of the existence of the Maxwell points not including the fundamental root.

\noindent
{\it Bifurcation set within ${\cal D}_c$ plane}.- 
In this section we consider only the critical point $v_0 = 0$. Then, $W_0 \rightarrow w_0(\alpha; \rho_2, \rho_1)$ and we can write Eqs.(\ref{pot-geom2}), and (\ref{alphaderivative}) as:
\begin{equation}
w_0(\alpha; \rho_2, \rho_1) = \rho_2 \,  p_A(\alpha) \, + \rho_1 \, p_B(\alpha) \, + p_0(\alpha) \, ,
\label{pot-geom3}
\end{equation}
\begin{equation}
U_\alpha =  \frac{\alpha}{q_1(\alpha)} \, w_1(\alpha; \rho_2, \rho_1) \, .
\label{alphaderivative2}
\end{equation}

We want to evaluate the bifurcation set of the $\alpha_c \neq 0$ critical points. These critical points are the roots of the equation
\begin{equation}
w_1(\alpha; \rho_2, \rho_1)\Big\vert_{\alpha_c} = 0 \, .
\label{alpha1derivative}
\end{equation}
As we mention in point (ii) of Sect. III, we begin by considering the mapping, ${\cal M}$, from the critical manifold to the essential parameter space
\begin{equation}
{\cal M}: \qquad \left\{\left(\alpha_c, \rho_2, \rho_1\right)\right\} \longrightarrow \left\{\left(\rho_2, \rho_1\right)\right\} \, 
\label{mapping}
\end{equation}
where $(\rho_2, \rho_1)$ represents the essential parameter space.

In order to find the singular points at the parameter space of this mapping, instead of solving for $\alpha_c$ in Eq.
(\ref{alpha1derivative}) we solve for $\rho_1$ and rename
\begin{equation}
\alpha_c \rightarrow \lambda_1 \, , \quad  \rho_2 \rightarrow \lambda_2 \, , \quad \rho_1  \rightarrow S_1(\lambda_1, \lambda_2) \, .
\label{renaming}
\end{equation}
The mapping is singular if its Jacobian determinant is zero, or equivalently, if
\begin{equation}
\frac{\partial S_1(\lambda_1,\lambda_2)}{\partial \lambda_1} = 0 \, .
\end{equation}

Finally, we solve for $\lambda_2$ in this last equation. This solution gives a critical $\lambda_2$ as a function of $\lambda_1$:
\begin{equation}
\lambda_{2c} \equiv {\cal S}_2(\lambda_1) \, .
\end{equation}

The mapping ${\cal M}$ in Eq.(\ref{mapping}) is invertible, except for a smooth curve on the critical manifold in the tri-dimensional space, given by: 
\begin{equation}
\left\{\left(\alpha_c, \rho_2, \rho_1\right)\right\} = (\lambda, {\cal S}_2(\lambda),{\cal S}_1(\lambda))
\end{equation}
where we have defined
\begin{equation}
{\cal S}_1(\lambda) \equiv S_1(\lambda, {\cal S}_2(\lambda) ) \, .
\end{equation}

The separatrix, ${\cal C}_1$, for the critical points $\alpha_c \neq 0$, is the set of points on the essential parameter space obtained parametrically as:
\begin{equation}
{\cal C}_1: \qquad \Big\{\Big({\cal S}_2(\lambda), {\cal S}_1(\lambda)\Big) \Big\vert \; \lambda \,  \epsilon \, {\cal R}e^+ \Big\}  \, .
\label{bifurcation1}
\end{equation}

This separatrix, ${\cal C}_1$, curve surges from the {\it germ} at the cusp point: 
\begin{equation}
\Big({\cal S}_2(0), {\cal S}_1(0)\Big) = (12/19, -4/3) \, ,
\end{equation}
as a numerical evaluation can demonstrate.

\noindent
{\it The Maxwell set  ${\cal C}_M$ within ${\cal D}_c$ plane}.- 
 In the first place, we observe that for a given point on the essential parameter space we obtain a particular graph of potential $\widetilde{V}$. Further, suppose that we have chosen a point belonging to the Maxwell set. Then, at this particular point $(\rho_2, \rho_1) $ the form of the graph is such that both the fundamental critical point $\alpha_0 = 0$ and another $\alpha_c \neq 0$ critical point satisfy the equality
 \begin{equation}
 \widetilde{V}(\alpha_0; \rho_2, \rho_1) = \widetilde{V}(\alpha_c; \rho_2, \rho_1) =0 \, .
 \end{equation}
 Also, it is clear from Eqs.(\ref{pot-geom1}), and (\ref{pot-geom3}), that the following equation is satisfied
\begin{equation}
w_0(\alpha; \rho_2, \rho_1)\Big\vert_{\alpha_c} = 0 \, .
\label{w0crroots}
\end{equation}

At the second place, we observe that in a more general $(\rho_2, \rho_1) $ point, a root $\alpha^{(j)} \neq 0$ of the equation:
\begin{equation}
w_0(\alpha; \rho_2, \rho_1) = 0  
\label{w0roots}
\end{equation}
means that the graph of potential $\widetilde{V}$ is crossing the horizontal $\alpha$-axis, at a distance $\alpha^{(j)}$-away from the origin. As the $(\rho_2, \rho_1) $ point changes towards a Maxwell point, two different roots,  $\alpha^{(j)}$ and $\alpha^{(k)}$, of this last equation coalesce. Obviously, this also implies that at the coalescence $\alpha^{(j)}$-value the  $\alpha$-horizontal axis becomes a tangent, and thus a critical Maxwell point will be found. From the facts pointed in this paragraph we propose a method, inspired in catastrophe theory, that produces the Maxwell set as the singularities on the parameter space  $(\rho_2, \rho_1) $ of a mapping from the three-dimensional surface, $(\alpha^{(j)}; \rho_2, \rho_1)$, of all different $\alpha^{(j)}$-root points of Eq. (\ref{w0roots}), obtained by a continuous variation of the essential parameters, similar to the singularity illustrated in Fig. (\ref{cuspide}).

The coalescence points of the roots $\alpha^{(j)} \neq 0$ of Eq.(\ref{w0roots}) can be found by considering the singularities of the mapping:
\begin{equation}
{\cal M}_{M}: \; \left\{\left(\mu_1,\, \mu_2,\, T_1(\mu_1,\mu_2)\right)\right\} \longrightarrow \left\{\left(\rho_2, \rho_1\right)\right\} \, .
\label{Maxmapping}
\end{equation}
$T_1(\mu_1,\mu_2)$ obtained solving for $\rho_1$ in Eq. 
(\ref{w0roots}) and being renamed as follows:
\begin{equation}
\alpha^{(j)}  \rightarrow \mu_1 \, , \quad  \rho_2 \rightarrow \mu_2 \, , \quad \rho_1  \rightarrow T_1(\mu_1, \mu_2) \, .
\label{Maxwellrenaming}
\end{equation}
The mapping is singular if its Jacobian determinant is zero, or equivalently, if
\begin{equation}
\frac{\partial \, T_1(\mu_1,\mu_2)}{\partial \mu_1} = 0 \, .
\end{equation}
We obtain the critical $\mu_2$, by solving it from this last equation as a function of $\mu_1$:
\begin{equation}
\mu_{2c} \equiv {\cal T}_2(\mu_1) \, .
\end{equation}

The separatrix, ${\cal C}_{M}$ for the  roots $\alpha^{(j)} \neq 0$ is the loci on the essential parameter space obtained as
\begin{equation}
{\cal C}_{M}: \; \Big\{\Big({\cal T}_2(\mu), {\cal T}_1(\mu)\Big) \Big\vert \; \mu \,  \epsilon \, {\cal R}e^+ \Big\} \, ;
\label{Maxwellset}
\end{equation}
where we have defined
\begin{equation}
{\cal T}_1(\mu) \equiv T_1(\mu, {\cal T}_2(\mu) ) \, .
\end{equation}
This Maxwell curve surges from the {\it germ} at the cusp point: 
\begin{equation}
\Big({\cal T}_2(0), {\cal T}_1(0)\Big) = (12/19, -4/3) \, ,
\end{equation}
as a numerical evaluation can demonstrate
\cite{mat} (see also Fig. (\ref{Separatrix})).

The parametric plotting on the essential parameter space of the bifurcation sets ${\cal C}_{0}$ (Eq. (\ref{separatrix0})), ${\cal C}_{1}$ (Eq. (\ref{bifurcation1})) and the Maxwell set ${\cal C}_{M}$ (Eq. (\ref{Maxwellset})), defines the Separatrix of the system. The character of each equilibrium point remains the same at each distinct region, and a sudden qualitative changes take place when this separatrix is crossed as a result of a general variation of the Hamiltonian parameters. These sudden changes are treated as a mathematical catastrophes and constitute the quantum phase transitions of the ground and the excited states for the cluster model Hamiltonian in Eq. (\ref{Ham}).

In Fig.(\ref{Separatriz3}) we choose a point at every region of the Separatrix, and at every component of this Separatrix. In Fig. (\ref{Grida}) 
to Fig. (\ref{Gridc})) we plot the corresponding potentials.

%%%%%%%%%%%%%%%%%%%%%%%%%%%%%%%%%%%%%%
\begin{figure}[htb] %  figure placement: here, top, bottom, or page
  \centering
  \includegraphics[width=3in]{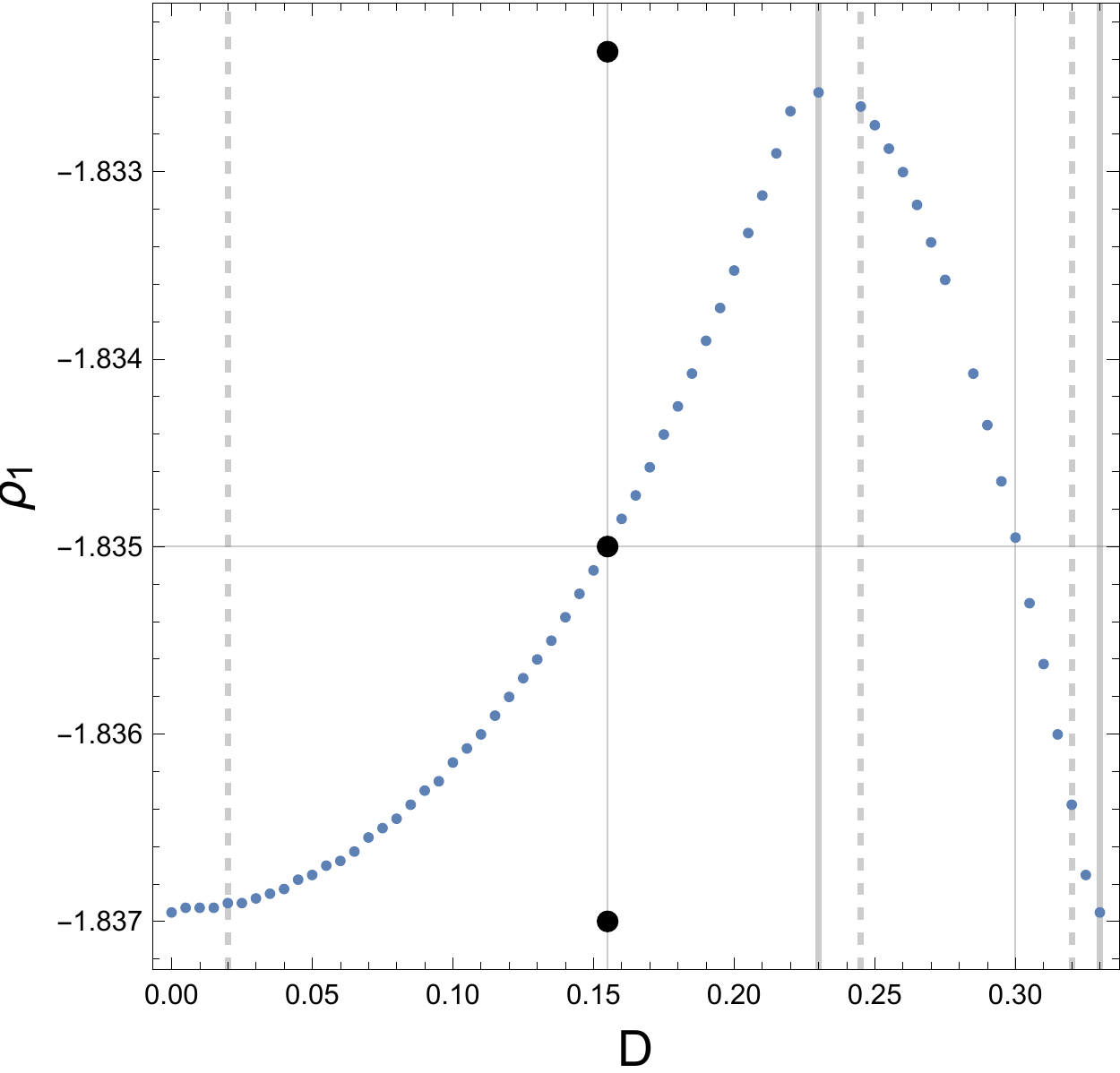} 
 \caption{
The Maxwell set within the $(D,\rho_1)$ sub-space, is plotted at  a constant
$\rho_2=$ 1.12324. The region of positive curvature ($D<0.21$) of
the Maxwell set corresponds to a {\it spherical-deformed} shape 
coexistence, while the negative curvature region ($D>0.21$) corresponds to 
a {\it deformed-deformed} shape coexistence. The horizontal
line at $\rho_1=$ -1.835 cuts the Maxwell set at two points in both 
regions, with the energy minimum going through a first order phase transition in each one,
as it is illustrated in Fig. (\ref{E1206m002}).
}
\label{MaxwDz60pointsbb}
\end{figure}
%%%%%%%%%%%%%%%%%%%%%%%%%%%%%%%%%%%%%
%%%%%%%%%%%%%%%%%%%%%%%%%%%%%%%%%%%%%

%%%%%%%%%%%%%%%%%%%%%%%%%%%%%%%%%%%%%%
\begin{figure}[t] %  figure placement: here, top, bottom, or page
   \centering
   \includegraphics[width=3.2in]{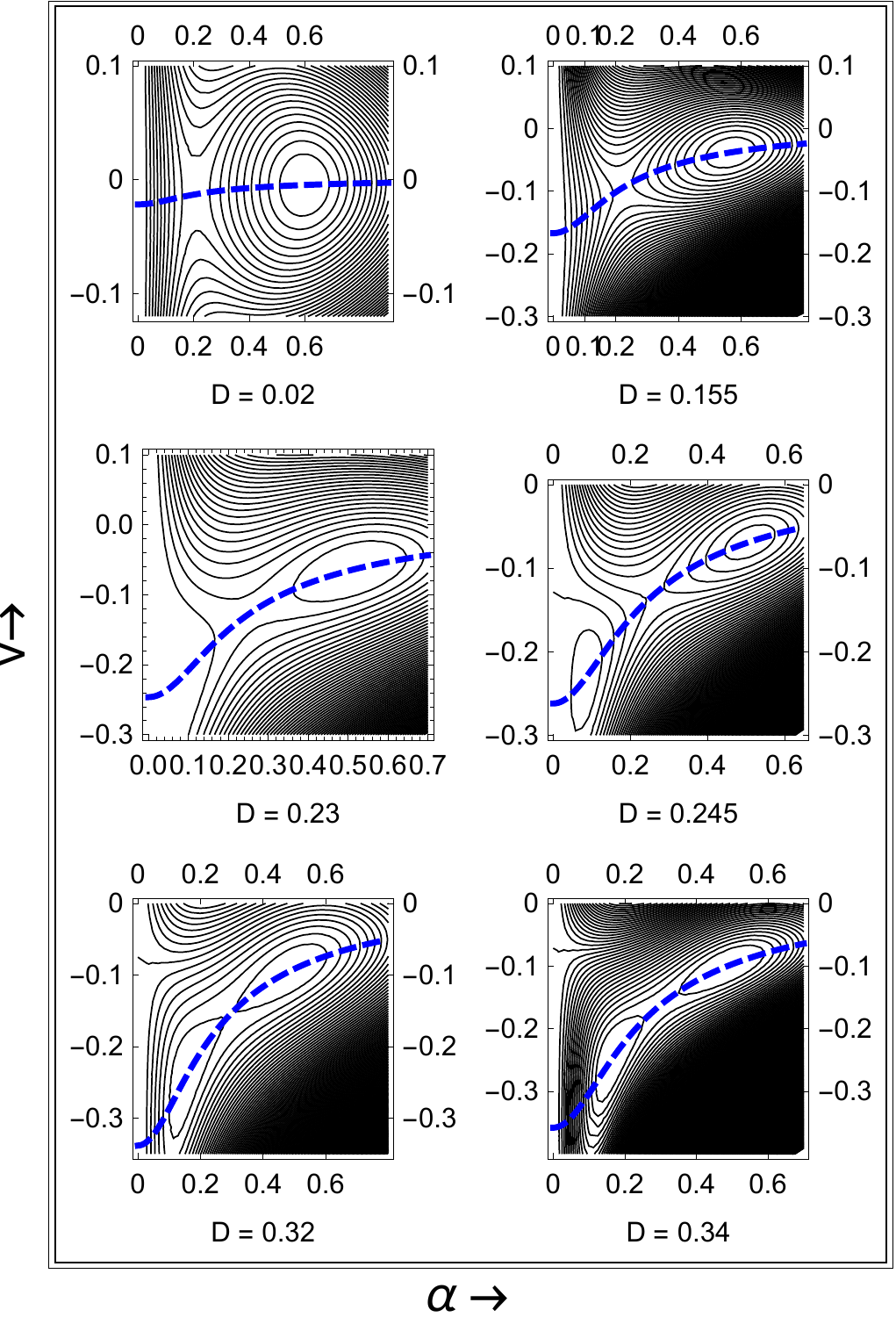} 
   \caption{
Curves of semi-classical energy levels for some points within the Maxwell set in Fig. (\ref{MaxwDz60pointsbb}). The corresponding $D$-parameter value is indicated. From $D=0.21$ on, 
both minima are outside the origin. A phase transition happened in
the coexistence of forms, namely from {\it spherical-deformed}
to {\it deformed-deformed}. 
	}
\label{Maxw60fff}
\end{figure}
%%%%%%%%%%%%%%%%%%%%%%%%%%%%%%%%%%%%%
%%%%%%%%%%%%%%%%%%%%%%%%%%%%%%%%%%%%%%

%%%%%%%%%%%%%%%%%%%%%%%%%%%%%%%%%%%%%%
\begin{figure}[t] %  figure placement: here, top, bottom, or page
  \centering
  \includegraphics[width=3.2in]{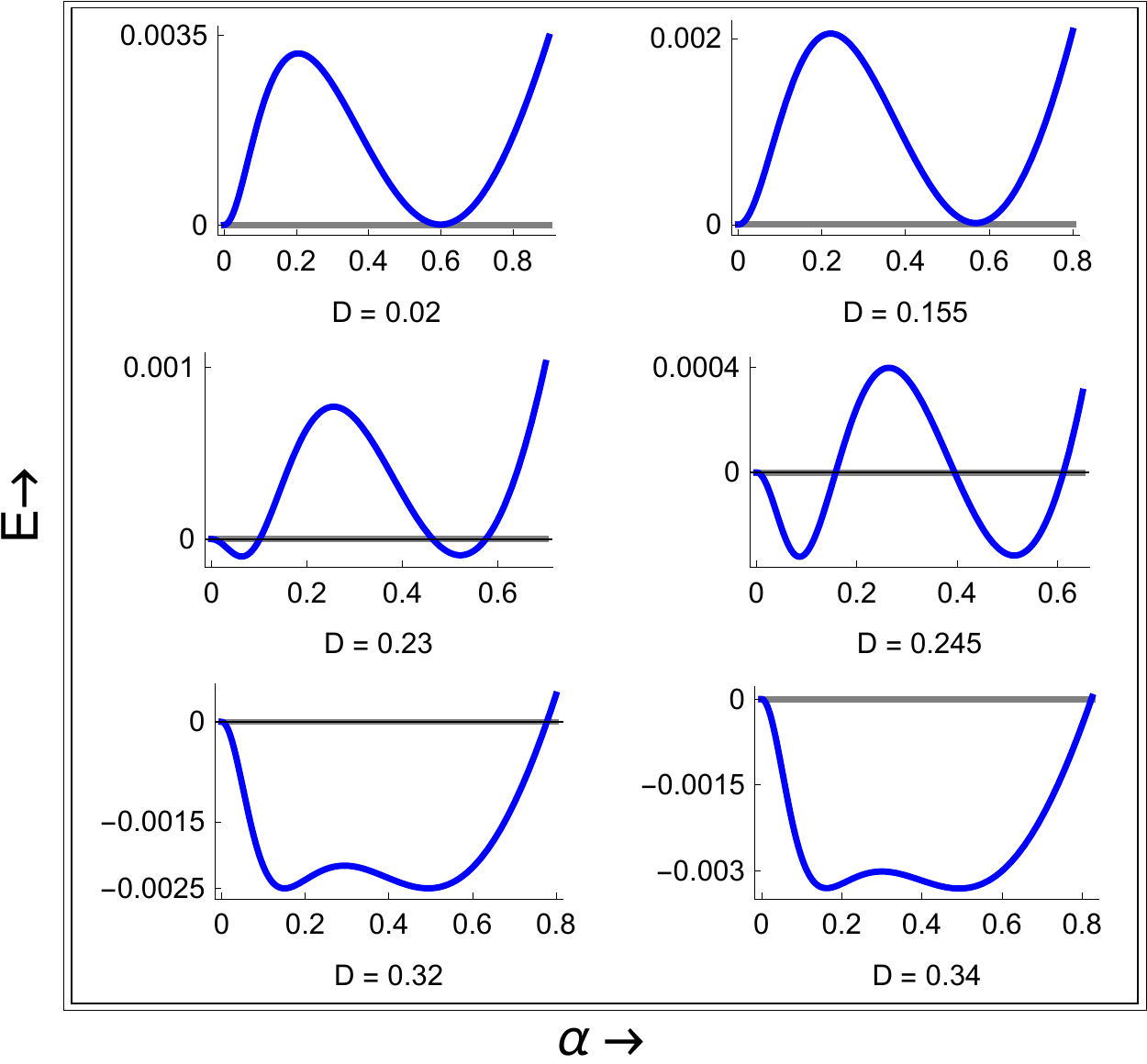} 
  \caption{
Curves of the semi-classical energy along the steepest descent curves given in Fig. 	
(\ref{Maxw60fff}).
}
\label{Eaxw60fff}
\end{figure}
%%%%%%%%%%%%%%%%%%%%%%%%%%%%%%%%%%%%%

%%%%%%%%%%%%%%%%%%%%%%%%%%%%%%%%%%%%%%
\begin{figure}[t] %  figure placement: here, top, bottom, or page
   \centering
   \includegraphics[width=3.2in]{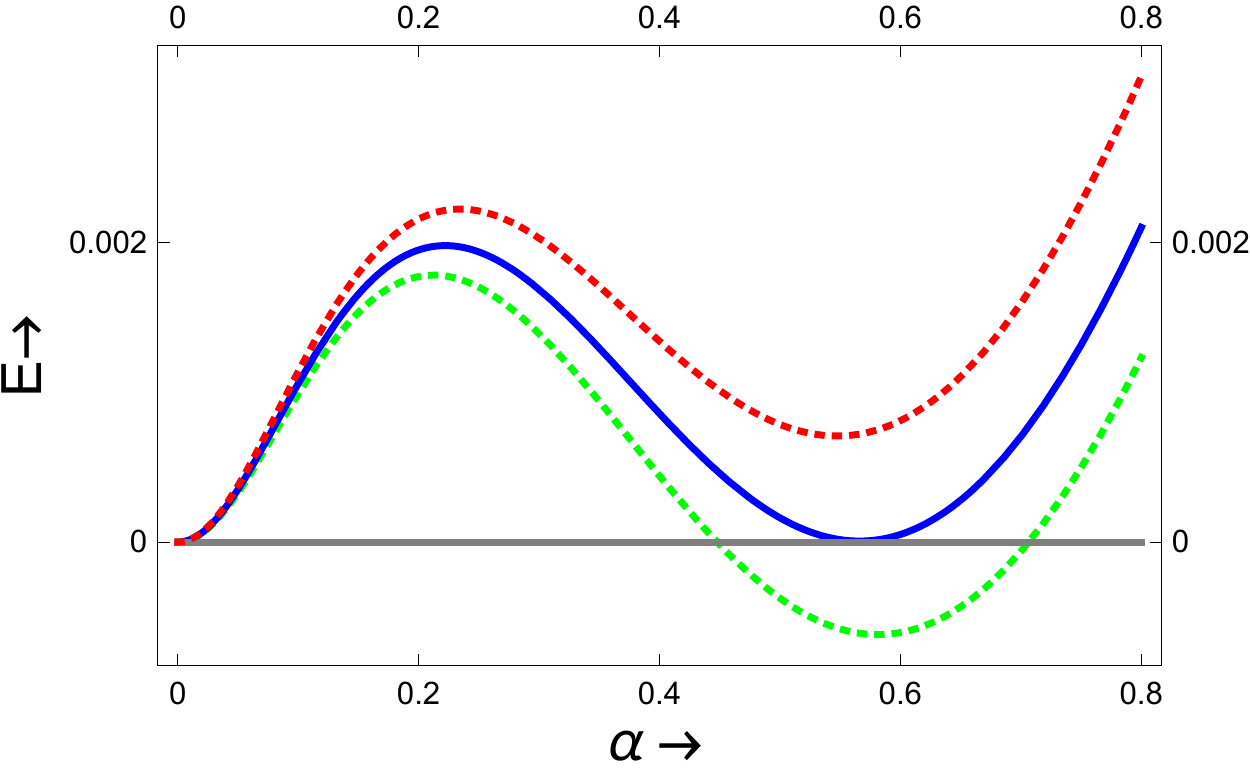} 
   \caption{
Plots of the semi-classical energy 
taken at the value $D=0.155$, corresponding to three bold points in Fig. (\ref{MaxwDz60pointsbb}) at the 
vertical line. By crossing the Maxwell set from above, nucleus experiences a shape-phase-transition; it goes from: ({\it spherical-minimum})\&({\it deformed-metastable}),   to a 
({\it deformed-minimum})\&({\it spherical-metastable}); with a shape coexistence in between at the Maxwell set.}
\label{EM06155uodv3.pdf}
\end{figure}
%%%%%%%%%%%%%%%%%%%%%%%%%%%%%%%%%%%%%
%%%%%%%%%%%%%%%%%%%%%%%%%%%%%%%%%%%%%%

%%%%%%%%%%%%%%%%%%%%%%%%%%%%%%%%%%%%%%
\begin{figure}[t] %  figure placement: here, top, bottom, or page
   \centering
   \includegraphics[width=3.25in]{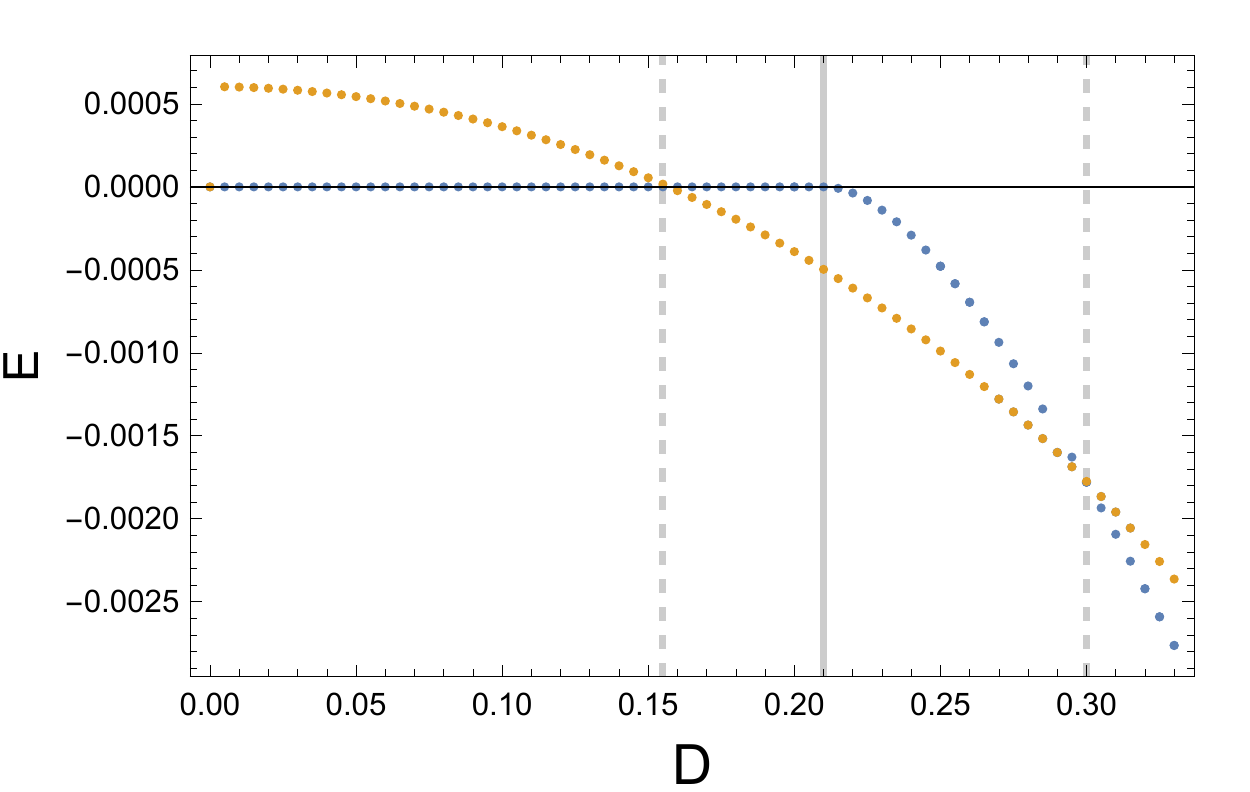} 
  \caption{
The absolute, as well as the secondary, minima vs. $D$, are 
plotted for a given pair of parameters: $(\rho_2,\rho_1)$ =
(1.12324,-1.835). This plot corresponds to the horizontal line in
Fig. (\ref{MaxwDz60pointsbb}). At the intersection with the Maxwell set,
the absolute energy minimum goes through a slope discontinuous change: a first order phase
transition. The secondary energy minimum presents at $D=0.21$ a further second order 
phase transition. At this point, the Maxwell set, in Fig. (\ref{MaxwDz60pointsbb}), changes its curvature from positive, to negative.
}
\label{E1206m002}
\end{figure}
%%%%%%%%%%%%%%%%%%%%%%%%%%%%%%%%%%%%%

\section{General Sub-parameter spaces, $D \neq 0$}
\label{sec:Dneq0}

In the general case there is a transition between
the $SU_R(3)$ and the $SO_R(4)$ dynamical symmetry and the Hamiltonian contains
interactions from both groups ($0 < x < 1$), thus given place to a more elaborated parameter competition, and to a richer shape-phase-transition behavior. Shape coexistence founds its roots at the Germ point ${\cal G}_0$ in Eq. (\ref{Germ0}). Within the critical plane ${\cal D}_c$ the Maxwell set ${\cal C}_M$ establishes shape coexistence and plays the role of a separatrix, dividing the critical ${\cal D}_c$ plane into quali\-ta\-tively different 
regions of behavior. In the present section we demonstrate that a more general shape coexistence stability can be attained outside this critical plane.

In order to investigate shape coexistence for a most general point, $(\rho_2, \, \rho_1, D)$, of the essential parameter space, it results convenient to take a fixed $\mu$ value into Eq. (\ref{Maxwellset}) and depart from the corresponding point of the Maxwell set ${\cal C}_M$, by increasing continuously the $D$ parameter value from $D=0$. In this fashion, within the $(D, \rho_1)$ sub-space, a section cut for the extended Maxwell set surface is plotted as the parametric curve (see Fig. (\ref{MaxwDz60pointsbb})):
\begin{equation}
\Big(D, \rho_1(D)\Big)\bigg\vert_\mu = \Big(D,\, {\cal T}_1(\mu) +\delta(D,\, \mu) \Big)\bigg\vert_\mu  \, .
\end{equation}
The $\delta(D,\, \mu)$ values were evaluated by designing a computer numerical program \cite{mat}
implemented for this purpose for any $\mu$ parameter chosen. 
In Fig. (\ref{MaxwDz60pointsbb}), the value $\mu= 0.6$ was taken. 
A collection of energy potential surfaces are presented in Fig. (\ref{Maxw60fff}) for various selected  points of this Maxwell set. The corresponding $D$ parameter value is explicitly given for every case, and in Fig. (\ref{MaxwDz60pointsbb}) those values are indicated by means of vertical lines.

The behavior of the semiclassical energy, as parameter $D$ evolves, is shown in Fig. (\ref{Eaxw60fff}) at some points of the Maxwell set; each graph follows a steepest descent path, indicated as a dashed curve in Fig. (\ref{Maxw60fff}).

Maxwell set in $D \neq 0$ regions, also governs stability behavior. The fundamental root $\alpha_0$ goes through a stable-metastable phase transition when Maxwell set is crossed from the above, as its is shown in Fig. (\ref{EM06155uodv3.pdf}).

The critical 
angular variables, $v_{cr}$, satisfy Eq. (\ref{vderivative}). From the expression in Eq. (\ref{pot-geom2}) it follows that outside  the ${\cal D}_c$ plane the angular critical values such that $v_{cr} \neq 0$.  These critical values are the roots of a fourth degree polynomial ${\cal P}(v)$

\begin{eqnarray}
{\cal P}(v) &=&   v^4 - r_1 \,  v^3 + ((r_1/2)^2 + r_2^2 - 1) \,  v^2 \\  \nonumber
   && + \,  r_1 \, v  - (r_1/2)^2;
 \end{eqnarray}
where $r_2$ and $r_1$ depend on the $\alpha$ variable and $C, \, D, \, \Omega$ parameters in the following way: 

\begin{equation}
r_2(\alpha, C,\Omega) =  \frac{ s(\alpha, \Omega)}{t(\alpha, C) } \, ;
\end{equation}

\begin{equation}
r_1(\alpha, C,D) = -2\, \frac{ \kappa(\alpha, D)}{t(\alpha, C) } \, .
\end{equation}
Here, $\kappa(\alpha, D)$, $s(\alpha, \Omega)$, $t(\alpha, C)$, are polynomials in even powers of  $\alpha$, positive integer coefficients and a non-zero constant term. Each one of these polynomials are respectively multiplied by one factor corresponding to the parameter indicated in its argument. The general behavior of these roots, its bifurcations and stability in the $(r_2,\, r_1)$ parameter space was studied in \cite{lopez2011}. 

One of these $v_{cr}$ describes all critical points of the $\widetilde{V}(\alpha, v)$ ener\-gy surface. This can be appreciated in 
Fig. (\ref{Maxw60fff}); where a thick dotted line goes through a steepest descend path obtained when $\widetilde{V}$ ener\-gy surface is restricted to the  $v_{cr}(\alpha, \rho_2(\mu), \rho_1(\mu), D)$ in its argument.

In Fig. (\ref{E1206m002}), there are two values for $D$ where the energy minima are degenerate: Where the Maxwell set is intersected.
In the first intersection a spherical-deformed coexistence appears
and in the second one a deformed-deformed coexistence. In Fig. (\ref{MaxwDz60pointsbb})
the Maxwell 
set within an arbitrary $(D,\rho_1)$ sub-space is presented: here we choose the 
$\rho_2=$ 1.12324 value. It is important to observe that there are two qualitatively different regions for this Maxwell set: A region of positive curvature ($D<0.21$) of
the Maxwell set corresponds to a {\it spherical-deformed} shape 
coexistence, while  a region of negative curvature ($D>0.21$) corresponds to 
a {\it deformed-deformed} shape coexistence. 

\begin{figure}[t] 
   \centering
   \includegraphics[width=3.25in]{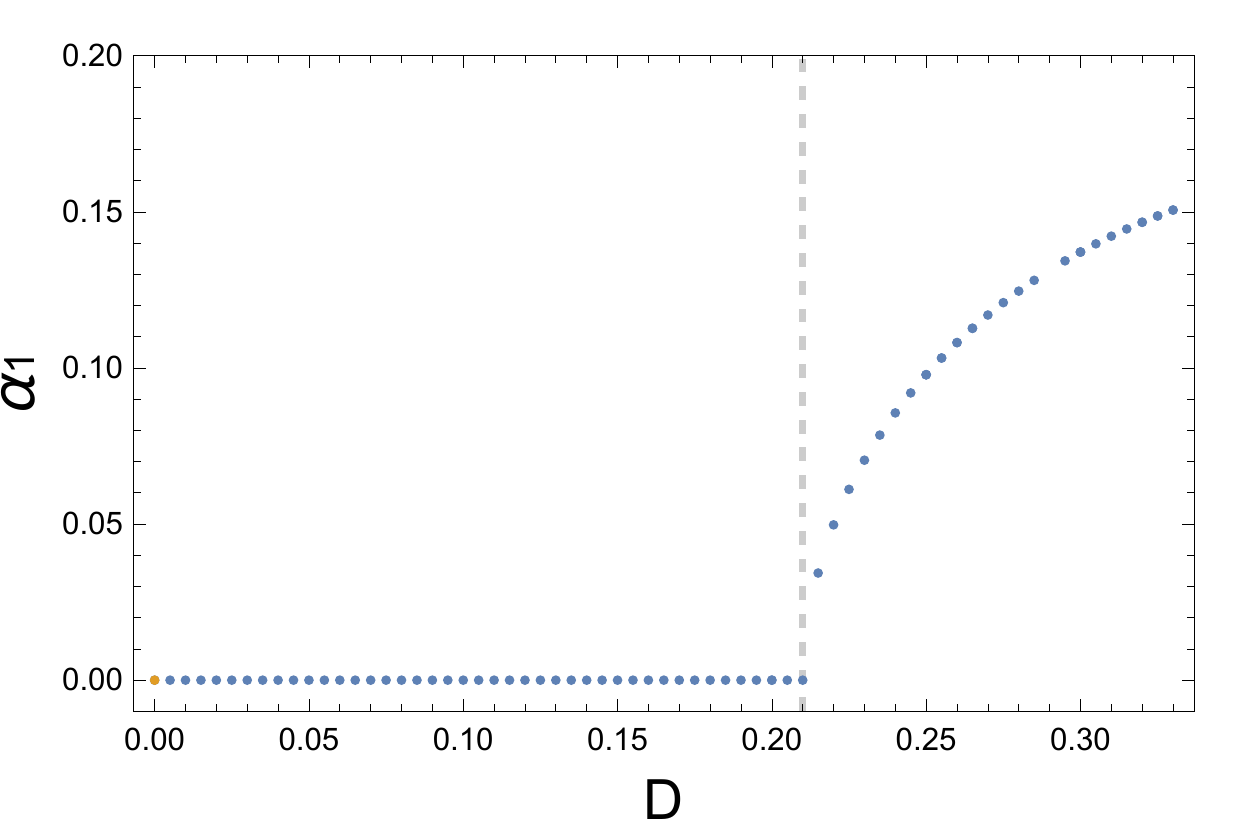} 
   \caption{The critical $\alpha_1$ deformation experiences a phase transition of first order at the value $D=0.21$; however, the corresponding critical energy value,  in Fig. (\ref{E1206m002}), experiences a second order phase transition there, since its slope remains continuous, but its curvature changes discontinuously.
}
\label{E1206m003}
\end{figure}

As a particular example with this $\rho_2=1,12324$ value, we choose a horizontal
line at $\rho_1=-1.835$ in Fig. (\ref{MaxwDz60pointsbb}); then this line cuts the Maxwell set at two points in both 
regions. Stability of the system is represented by the energy {\it absolute} minimum curve; this curve experiences a slope discontinuity at two points when parameter $D$ evolves, that is, its critical energy goes through a first order phase transition at each intersection. Secondary critical energy minimum,  experiences a further, second order, shape transition; this happens at the same $D$ point when the Maxwell set goes through a point of inflection; however, it is important to observe that the corresponding shape deformation of this metastable energy goes through a first order shape transition, as it is calculated, and  illustrated at Fig. (\ref{E1206m003}).

Finally, for any arbitrary parameter $\rho_2$, by means of  a similar numerical calculation, as in the past example, we can study how the system qualitative behavior evolves; then,  the Maxwell set will span a surface separatrix in the entire essential parameter space.

%%%%%%%%%%%%%%%%%%%%%%%%%%%%%%%%%%%%%%
%%%%%%%%%%%%%%%%%%%%%%%%%%%%%%%%%%%%%

\bigskip
\section{Conclusions}
\label{sec:Conclusions}

In this contribution we applied the catastrophe theory to a non-trivial and sufficiently complex model, namely the SACM, which shows the usefulness of the catastrophe theory and also its methods, on how to apply it in further complex models; not necessarily restricted to nuclear physics \cite{elmgrether}.

By means of coherent states and the catastrophe theory, the separatrix of an algebraic cluster model was constructed. We have discussed the salient features that allow to discriminate between a region in which phase transitions appear and a region that exhibits coexisting shapes of this model. We established the germ point on the essential parameter space. When evaluated at this  {\it germ} in the parameter space, the potential becomes: a) six-fold degenerated at its minimum, b) parameter independent and c) a ratio of two polynomials. 
At its  {\it germ} point the potential $\widetilde{V}$ has no other minimum critical point, and its maximum is attained only at infinity. If any of the parameters vary, going out of this {\it germ}, then other shapes and phase-shapes-transitions can occur. The {\it germ} is the locus of the separatrix cusp point, and by crossing this point we observe a 
{\it sixth order phase transition}, 
the highest one in this model. Here we limit our attention to regions very near the $SU(3)$ limit on the model, where the fundamental root of the potential play a central role, though, it was also shown on how to extend it to larger
deviations. 

In a future work it remains to consider equilibrium  minimal points far from the origin, in order to extend the Maxwell set of shape coexistence forms.

\section*{Acknowledgments}
P. O. H. acknowledges financial support from DGAPA (PAPIIT IN 100418)
and CONACyT (grant 251817).

%%%%%%%%%%%%%%%%%%%%%%%%%%%%%%%%%%%%%

\bigskip

\end{document}